# Spin-orbit-torque driven magnetoimpedance in Pt-layer/magnetic-ribbon heterostructures

M. R. Hajiali[1,†], S. Morteza Mohseni[2,†,&], L. Jamilpanah[2], M. Hamdi[2],
S. E. Roozmeh[1], S. Majid. Mohseni [2,*]

[1]*Department of Physics, University of Kashan, 87317 Kashan, Iran*
[2]*Faculty of Physics, Shahid Beheshti University, Evin, 19839 Tehran, Iran*

When a flow of electron passes through a paramagnetic layer with strong spin-orbit-coupling such as platinum (Pt), a net spin current is produced via spin Hall effect (SHE). This spin current can exert a torque on the magnetization of an adjacent ferromagnetic layer which can be probed via magnetization dynamic response, e.g. spin-torque ferromagnetic resonance (ST-FMR). Nevertheless, that effect in lower frequency magnetization dynamic regime (MHz) where skin effect occurs in high permeability ferromagnetic conductors namely the magneto-impedance (MI) effect can be fundamentally important which has not been studied so far. Here, by utilizing the MI effect in magnetic-ribbon/Pt heterostructure with high transvers magnetic permeability that allows the ac current effectively confined at the skin depth of ~100 nm thickness, the effect of spin-orbit-torque (SOT) induced by the SHE probed via MI measurement is investigated. We observed a systematic MI frequency shift that increases by increasing the applied current amplitude and thickness of the Pt layer (varying from 0 nm to 20 nm). In addition, the role of Pt layer in ribbon/Pt heterostructure is evaluated with ferromagnetic resonance (FMR) effect representing standard Gilbert damping increase as the result of presence of the SHE. Our results unveil the role of SOT in dynamic control of the transverse magnetic permeability probed with impedance spectroscopy as useful and valuable technique for detection of future SHE devices.

[*]Corresponding author's email address: m-mohseni@sbu.ac.ir, majidmohseni@gmail.com
† These authors contributed equally.
& Current affiliation: Fachbereich Physik and Landesforschungszentrum OPTIMAS, Technische Universität Kaiserslautern, 67663 Kaiserslautern, Germany.

Spin-orbit torques (SOTs) generated by current injection in ferromagnet (FM)/heavy-metal (HM) heterostructures have attracted considerable attention as a method to effectively manipulate the magnetization of thin FM films[1–7]. The spin Hall effect (SHE)[8] is reported to be the dominant source of the damping-like (DL) SOT in such heterostructures that is responsible for magnetization switching[9,10], domain wall (DW) motion[11–13], skyrmion manipulation[14,15] and high-frequency magnetization dynamics[16–18]. Rashba-Edelstein effect (REE) as another source of SOT is also present in FM/HM bilayers and depends on the interface structure of these bilayer and their corresponding thicknesses[19]. This mechanism can result in the exertion of field-like (FL) torque on FM.

Quantification of the SOTs in FM/HM heterostructures are based on spin-torque ferromagnetic resonance (ST-FMR)[20], planar Hall effect[21], low-frequency (~maximum to few 100 Hz) harmonic Hall voltage[6,22], spin Hall magnetoresistance[23–25], DW creep velocity[26] and magneto-optical effect[27]. They requires high frequency (few GHz) instruments or need a complicated assessment process, hence, demonstration of SOT materials and techniques in low frequency regime (MHz) via easy experimental process is desirable.

The studied heterostructure in this letter is made of a FM $Co_{68.15}Fe_{4.35}Si_{12.5}B_{15}$ ribbon and a thin layer of platinum (Pt). Such ribbon among soft magnetic material is one of the most promising candidates for the MI effect[28-30], primarily because of its application in low-cost and high sensitive magnetic sensors[31,32] and MI magnetic random access memory (MRAM)[33]. This effect causes a change in electrical impedance of a conducting FM with high transverse magnetic permeability ($\mu_t$) in the presence of a static magnetic field[34]. By applying an external magnetic field, the skin depth ($\delta$) changes due to change in $\mu_t$, thus varying the impedance of the FM. In the case of the ribbon, with width $l$ and length $L$, the impedance is approximately[35]

$$Z = (1-i)\frac{\rho L}{2l\delta} = \frac{(1-i)L}{2l}(\pi \rho f \mu_t)^{\frac{1}{2}} \qquad (1)$$

where $\rho$ is electric resistivity, $f$ is frequency of the current and $i$ = imaginary unit. Therefore, the impedance of the ribbon is a function of frequency, driving current and the external dc magnetic field ($H_{dc}$) through $\mu_t$ and $\delta$.

Here, we study SOT effect on the MI response of the Co-based amorphous ribbon (~30 µm)/Pt (0-20 nm) heterostructure by measuring its external magnetic field and frequency dependence impedance response. It is however noted that, the MI will be studied in a system including a thick FM layer, but based on the aforementioned skin effect, the current distributes at approximately 100 nm thickness close to the thin Pt layer deposited at the interface. This enables to uncover the dynamic of domain and DW in the present of SOT within a thin region of skin depth. We observe that the impedance response is strongly dependent on the thicknesses of the Pt layer and the applied current amplitude. Our results reveal the possibility of SOT detection in a FM using the impedance spectroscopy of FM in low frequencies ~MHz in high transvers magnetic permeability structures. Moreover, we have used ferromagnetic resonance (FMR) (see section S3 of the supplementary materials) for a better understanding of the mechanism happening in this system to represent the role of SOT effect based on fundamental and standard measurement to confirm the validity of our technique.

Amorphous Co-based ribbons (2 mm width, 30 mm length and ~30 μm thickness) were prepared by a conventional melt-spinning technique. Before deposition of Pt layer, about 40 nm of ribbons surface were sputter etched via Ar to have clean and oxygen free components. Pt thin layers with thickness of 10 nm and 20 nm were deposited on the soft surface (wheel side) of those ribbons in the Ar with gas pressure of 5 mTorr, base pressure better than $5\times10^{-8}$ Torr and growth rate of 3 nm/minute. (See supplementary materials including MI measurement, X-ray diffraction (XRD) analysis and FMR measurements)

Schematic illustration of a FM/HM heterostructure system and the definition of the Cartesian coordinate system in this work are presented in Fig. 1. The high $\mu_t$ of these ribbons allows the skin effect to occur in the MHz frequency range with thickness <100 nm. As shown schematically in Fig. 1(a), an ac charge current in the HM layer generates a pure spin-current, oscillating at the same frequency, perpendicular to the charge current direction thanks to the SHE. This oscillating spin current flows into the adjacent FM layer, exerts two different types of oscillating SOTs[5,6,36].

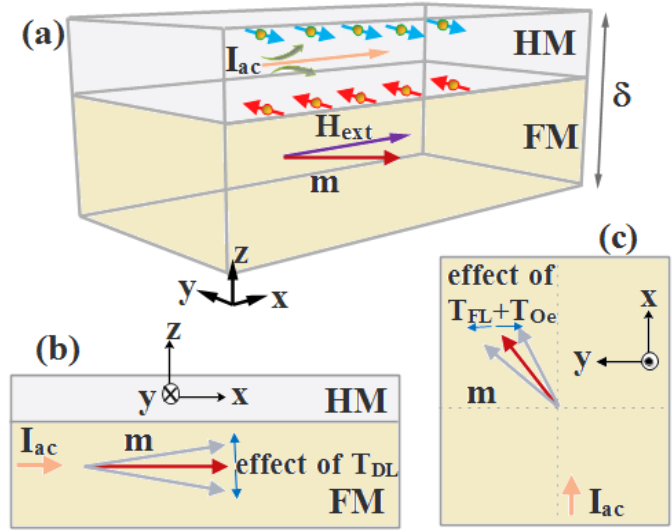

FIG 1: (a) Schematic illustration of a FM/HM heterostructure system. An in-plane charge current $I_{ac}$ generates a perpendicular spin current, which in turn generates SOTs acting on ferromagnetic moments. Oscillations of the magnetization due to (b) damping-like SOT ($T_{AD}$) and (c) the field-like SOT and Oersted field ($T_{FL} + T_{Oe}$) induced by an ac current. We should note here that this scenario happens at the skin depth $\delta$ of the ribbon.

They are field-like (FL) torque $\mathbf{T}_{FL} \sim \mathbf{m} \times \mathbf{y}$ and damping-like (DL) torque $\mathbf{T}_{DL} \sim \mathbf{m} \times (\mathbf{y} \times \mathbf{m})$, where $\mathbf{m}$ is the magnetization unit vector and $\mathbf{y}$ is the in-plane axis perpendicular to current flow direction $\mathbf{x}$ (Fig. 1 (b, c)). $\mathbf{T}_{DL}$ originates from the SHE in the adjacent HM layer. The magnitude of this $\mathbf{T}_{DL}$ depends on the transmission of spin current across the FM/HM interface[36,37]. $\mathbf{T}_{FL}$ can be originated from the REE at the FM/HM interface due to the structural inversion asymmetry or from the spin current through HM via the SHE[19]. When the magnetization lies in-the-plane of the bilayer sample, the action of $\mathbf{T}_{FL}$ is equivalent to an in-plane field $\mathbf{h}_{FL} \sim \mathbf{y}$, and that of $\mathbf{T}_{DL}$ establishes an out-of-plane field $\mathbf{h}_{DL} \sim \mathbf{m} \times \mathbf{y}$. Although we have a thick FM layer, at the studied frequency range (1-25 MHz) in our samples, the current passes through the skin depth $\delta$ of ribbon (~few nm to 100 nm), therefore the above mentioned scenario is valid in the MI measurement that we introduce here. The magnitude of $\mathbf{T}_{FL}$ varies significantly with thickness of FM layer[5], the type of FM and HM[38,39] and the direction of magnetization in the FM[37]. Two origins of $\mathbf{T}_{FL}$ are known generally, one due to REE and the other due to SHE. It is admitted that $\mathbf{T}_{FL}$ due to REE, reveals in FM/HM heterostructures with 1-nm-thick FM[41,42] and $\mathbf{T}_{FL}$ due to SHE remains very weak in metallic systems[43]. Also it is shown that very large $\mathbf{T}_{FL}$ occurs in magnetic tunnel junctions[44] and HM/nonmagnetic/FM/oxide heterostructures[45]. Therefore, in our studied heterostructure the contributions of $\mathbf{T}_{FL}$ can be neglected because of the metallic nature of layers and large thicknesses of FM layer. From now on, we consider both $\mathbf{T}_{DL}$ that comes primarily from the SHE and Oersted torque ($\mathbf{T}_{Oe}$) due to Oersted field generated from the charge current that depends upon the conductivity of each layer and skin depth $\delta$. As the thickness of FM layer is much larger than its skin depth, the Oersted field from FM layer can be important[20]. In order to detect the effect of these torques we carry out impedance measurement by applying an ac charge current with frequency $f$ to the samples, and investigate how the impedance $Z$ of the bilayer changes as a function of frequency and field.

Comparison of frequency sweep MI measurement for 0, 10 and 20 nm Pt is shown in Fig 2(a) where an external field of 120 Oe was applied to saturate the sample in the plane and an ac current with peak to peak amplitude of 66 mA was used to excite the sample. According to equation 1 and based on literatures arguments the impedance depends on current frequency $f$ and transverse magnetic permeability $\mu_t(f)$ with decreasing trend at high frequencies[46]. Therefore, with increasing $f$, the impedance of sample increases up to some frequency and further increase of $f$ results in the reduction of impedance where strong reduction of $\mu_t(f)$ occurs.

When an ac current flows through a FM layer the magnetization oscillates about its equilibrium position, y direction, due to Oersted field. Because of the presence of Pt layer, generated spin currents due to the SHE from Pt layer consequences into $T_{DL}$ that derives the magnetization oscillation in the *z* direction. It can be seen that the frequency of the maximum impedance of the sample shifts towards high-frequency values, increased from 17 MHz for 0 nm Pt to 18.5 and 19.5 MHz for 10 and 20 nm Pt deposited ribbons, respectively. There is another confirmation for this fact (that will be discussed later) that MI versus field shows reduced transverse anisotropy as the magnetization oscillation changed its orientation toward z direction. We speculate that the angle of precession decreases from that transversal orientation (without $T_{DL}$) and the peak position that represents the maximum $\mu_t$ goes to higher values as shown in Fig. 2(b). The magnitude of h$_{DL}$ can be affected by changing the thickness of the HM and FM[5] while we have varying thickness of the HM layer. The spin current in FM/HM heterostructures obeys $J_s(t) \approx 1 - \text{sech}(t_{HM}/\lambda_{Sd})$[20], where $t_{HM}$ is the thickness of HM and $\lambda_{Sd}$ is the spin-diffusion length. In this relation, as the spin diffusion length of Pt layers is in the Co$_{75}$Fe$_{25}$/Pt bilayer film was estimated to be 2.1± 0.2 nm[47], therefore $J_s(t)$ does not have to change for 10 and 20 nm thickness of Pt contrary to the frequency shift observed from Pt (10 nm) to that for Pt (20 nm), represented in Fig. 2(b). However, this effect can be explained based on the resistivity of FM and HM layers. The resistivity of Pt layer is $\rho$ =20 $\mu\Omega$ cm and that for the ribbon is $\rho$ =130 $\mu\Omega$ cm which might pinpoint as a fact that at the skin depth region the current in the thicker Pt deposited layer is more than when Pt thickness is 10 nm. This implies a fact that the current effect and therefore the SHE effect is more significant for sample deposited with 20 nm Pt with enhanced T$_{DL}$ that results in frequency shift.

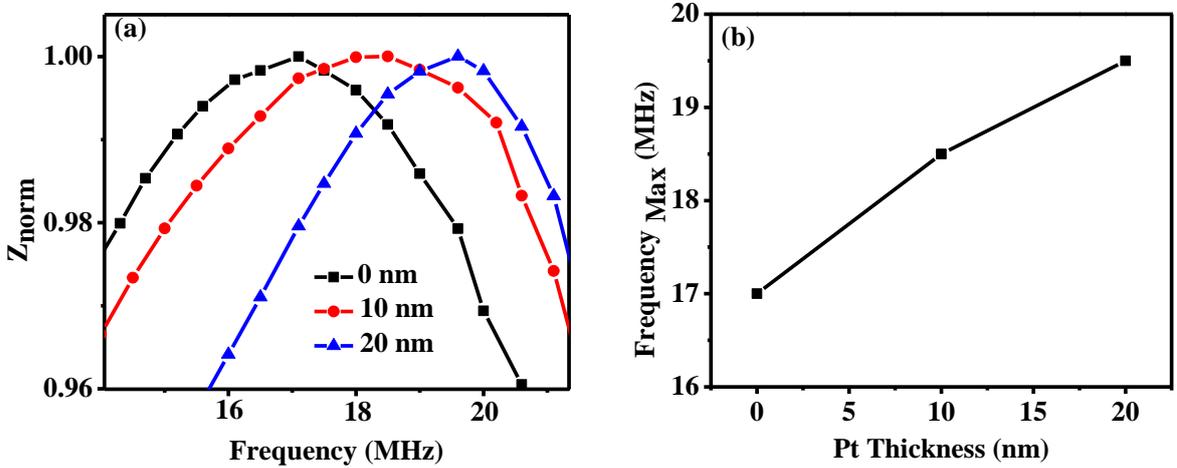

FIG. 2. (a) Frequency sweep of impedance measurement of 0, 10 and 20 nm Pt deposited ribbon normalized to maximum showing a shift towards higher *f*. (b) The maximum frequency vs Pt thickness obtained from (a) showing higher shift for higher thickness of Pt.

We measured the frequency sweep of the MI response with different amplitude of current applied to the samples to better elucidate the origin of the frequency shift. Considering the relation between the spin current $J_s$ and the charge current ($J_c$), increasing the applied current amplitude results in higher spin current generation and higher h$_{DL}$ magnitude (h$_{DL}$ $\propto J_s$)[43,48]. Frequency sweep impedance measurement against ac current with peak to peak amplitude of 33, 66 and 99 mA are shown in Fig. 3 (a-c) for 0, 10 and 20 nm Pt deposited ribbons while the ribbons were saturated at 120 Oe. It is clear from Fig. 3(a) that increasing the magnitude of the applied current for 0 nm Pt does not affect the peak position of the impedance. Whereas for 10 and 20 nm Pt deposited ribbons, increasing the amplitude of the applied current results in a shift in the maximum impedance frequency. A comparison between the maximum impedance frequency shift and the applied current for all samples is shown in Fig. 3(d). As can be seen, the role of current for 20 nm Pt is more pronounced with larger frequency-current slop.

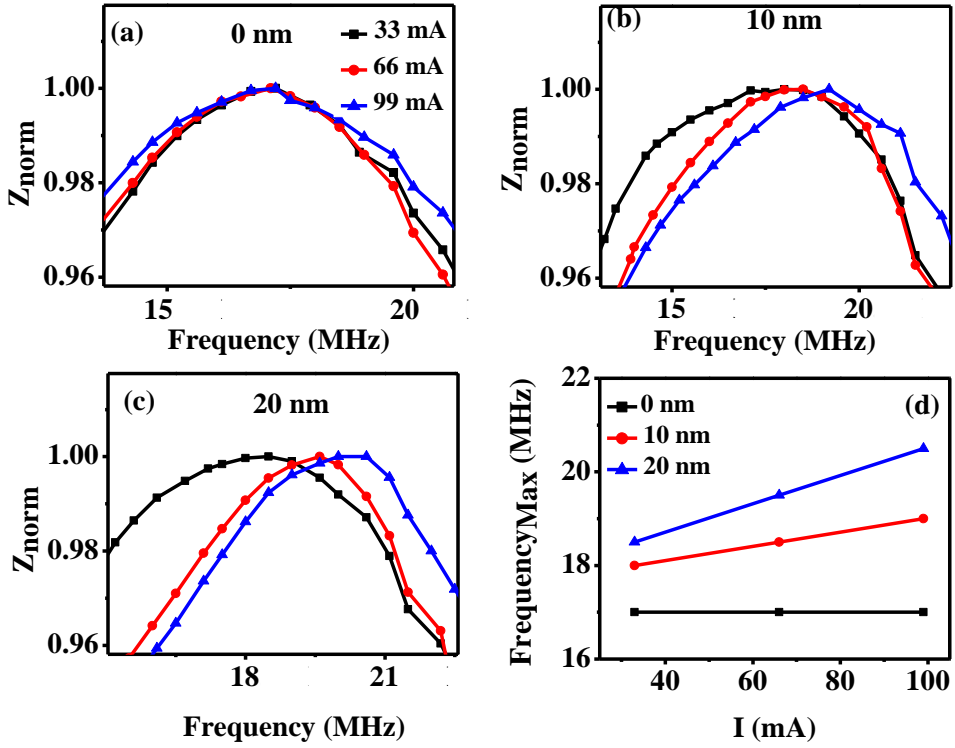

FIG. 3. Frequency sweep impedance measurement of (a) 0, (b) 10 and (c) 20 nm Pt at the presence of an external field of 120 Oe in different ac current peak to peak amplitude of 33, 66 and 99 mA with a higher frequency shift for higher driving currents. (d) Maximum frequency obtained from (a), (b) and (c) versus ac current amplitude indicating higher slope of increment for 20 nm Pt deposited ribbon than that for 10 nm Pt deposited one.

MI response of a ribbon can give us detailed information about magnetic anisotropy and transverse magnetic permeability. Therefore, we measured field sweep impedance measurement in an arbitrary frequency of 6 MHz for 0, 10 and 20 nm Pt samples with 66 mA current applied to the samples, as presented in Fig. 4(a, b). It is considered that based on equation S1 in supplementary materials, the MI decreases from 191% for bare sample to 169% for 10 nm Pt and 152% for 20 nm Pt deposited samples. This behavior is consistent with the $T_{DL}$ tends to reduce the $\mu_t$ by exerting a torque perpendicular to equilibrium angle of magnetization. As can be seen in Fig. 4(b), the bare ribbon shows a double peak behavior and Pt deposited ribbons show a single peak behavior. The observed single- or double-peak behaviors are associated with the longitudinal or transverse magnetic anisotropy with respect to the external field direction[49,50]. The disappearance of the transverse anisotropy in ribbon/Pt heterostructures could stem from the $T_{DL}$ which is perpendicular to the plane of the ribbon thus forces the magnetization from transverse alignment and reduces the transverse magnetic permeability. Furthermore, as another testifier, we repeated the experiment for ribbon sample coated with 20 nm Copper (Cu) coated layer and observed double peak behavior and no frequency shift similar to the bare ribbon (see FIG. S3 and S4 in supplementary materials). Cu is a representative light metal with weak spin–orbit coupling[51] and we expect to see double peak behavior.

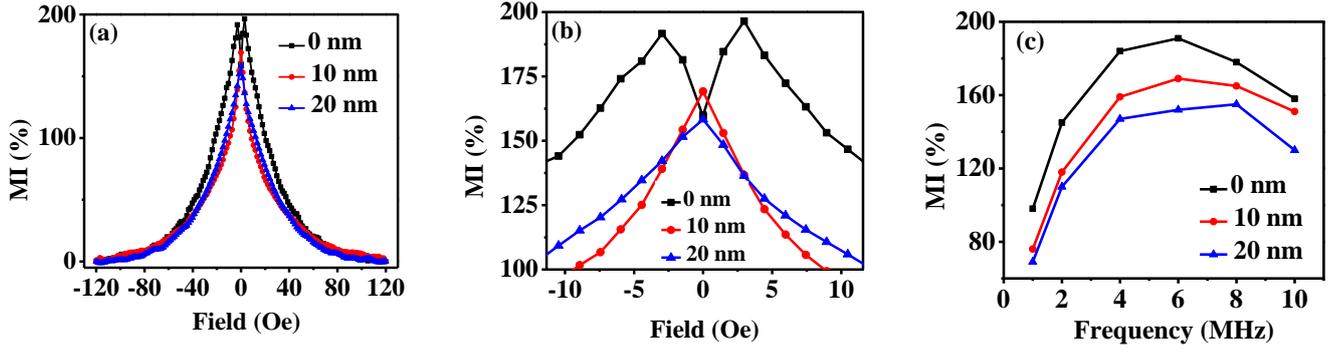

FIG. 4. (a) Comparison of the MI response of the 0, 10 and 20 nm Pt coated samples in an arbitrary frequency of 6 MHz and applied current of 66 mA. (b) A zoom-in window of MI in the low fields shows the disappearance of the double peak behavior for Pt deposited ribbons. (c) MI ratio of 0, 10 and 20 nm Pt deposited ribbons versus frequency of the applied current with a reductive behavior for Pt deposited ribbons.

The maximum MI ratio of all samples versus frequency are plotted in Fig. 4(c) to better illustrate the effect of Pt layer. MI measurements were done at different frequencies ranging from 1 MHz to 10 MHz. It is noted in Fig. 4(c) that for all investigated samples, with increasing frequency, the maximum MI ratio first increases, reaches to a maximum at a particular frequency (6 MHz), and then decreases for higher frequencies. This trend can be interpreted by considering the relative contributions of DW motion and moment rotation to the transverse magnetic permeability and hence to the MI[52,53]. Noted that as frequency increases well above 100 kHz, the contribution of DW motion is damped due to presence of the eddy current and moment rotation becomes dominant[46,53,54]. Thus, the MI ratio decreases at high frequencies. Here, the $\mu_t$ decreases, thus resulting in the observed drops of the MI ratio at all frequencies[46,52]. It is known that DW motion speed increases in the present of SHE[12,55,56]. MI ratio frequency peak is correlated to the DW relaxation and suggests how DW follows the ac current frequency or correlated with DW speed. Such increase in frequency for 20 nm Pt has same fashion as DW does in the present of SHE, dictating another qualitative confirmation.

In summery we have proposed that impedance spectroscopy can be used for detection of SOT resulting from the SHE in magnetic-ribbon /Pt heterostructures. Tunable impedance response correlated to SOT induced moment realignments within FM can be detected. We showed that in a magnetic-ribbon /Pt heterostructure, the acting $T_{DL}$ on FM changes not only the response of the MI of the system, but also tends to play with the transverse anisotropy of the magnetization that was probed as frequency shift in MI effect. Our results can open a practical route to study and understand the SHE probed via the MI effect in FM/HM heterostructures.

See supplementary material for a complete understanding of the detailed procedure of related experiments for the observation of SOT, including FMR measurements for magnetic-ribbon/Pt and MI for Cu sputtered ribbon.

S.M.M. acknowledges support from the Iran Science Elites Federation (ISEF).